# Computational Study of Magnetic Behaviour in Ni-Adsorbed $Nb_2C$-OF MXene using Density Functional Theory

Zarah Khan,[1] Saleem Ayaz Khan,[2] Ayesha Zaheer,[3] and Syed Rizwan[1*]

[1] Physics Characterization and Simulations Lab (PCSL), Department of Physics, School of Natural Sciences (SNS), National University of Sciences and Technology (NUST), Islamabad 44000, Pakistan

[2] New Technologies Research Centre, University of West Bohemia, Univerzitni 2732, 306 14 Pilsen, Czech Republic

[3] Department of Physics "Ettore Pancini", University of Naples Federico II, Piazzale Tecchio, 80, 80125 Naples, Italy

Corresponding author's: **Syed Rizwan** ; syedrizwan@sns.nust.edu.pk, syedrizwanh83@gmail.com
**Saleem Ayaz Khan**; sayaz_usb@yahoo.com

## Abstract

Magnetic 2D materials have achieved significantly consideration owing to their encouraging applications. A variation of these 2D materials by occurrence of defects, by the transition-metal doping or adsorption or by the surface functionalization can initiate both the spin-polarization and magnetic properties in these materials. Density functional theory (DFT) is used to determine the electric, magnetic properties along with the electronic structures and stability of synthesized two-dimensional materials. This work describes the magnetic properties of Ni-ad-$Nb_2C$-OF MXene. The study focuses on the computational approach based first principal calculation providing insight onto the magnetic properties of adsorbed compound and comparing it with pristine $Nb_2C$-OF MXene. The pristine $Nb_2C$-OF and Ni-ad-$Nb_2C$-OF structures are simulated and optimized using Wien2k software. Using exchange-correlational functionals; spin-GGA and spin-GGA+U (for Nickel U= 6eV), Ni-ad-$Nb_2C$-OF electronic band structure is found to be metallic having magnetic moment calculated +1.01516$\mu_\beta$ showing its non-superconducting and ferromagnetic ~~behaviour~~ properties. Owing to this magnetic ~~nature~~ properties, this 2D compound holds potential for emerging applications in spintronics and nanoscale magnetic data storage technologies.

**Key words:** Density Functional Theory (DFT), Total density of states (TDOS), magnetic moment, Ferromagnetism, Ni-adsorbed-$Nb_2C$-OF MXene.

## Introduction

Two dimensional materials are an emerging family of nano structured small scale dimensional materials and are believed as an encouraging candidate for various applications just as food, environment, energy storage and electronic devices owing to their unique electronic structures and large surface areas [1]. The accomplishment of two-dimensional (2D) layered materials magnetism has become a vital objective of scholars, because the 2D magnetic materials are not simply the basis for spintronics devices, but also the platform for understanding new physical phenomena. Commonly, magnetism can be introduced or increased in the non-magnetic and magnetic materials by doping magnetic atoms or using the near-neighbour effect of the interface [2]. Thus, the synthesis, properties, and applications of such novel 2D materials have now become an exhilarating area of interest in science and technology. Recently, it has been discovered that by using a sequence of chemical exfoliation and sonication, 2D layered materials can be prepared in bulk from three-dimensional (3D) layered host compound [3,4].

MXenes are one of such 2D materials. It is a new class of two-dimensional (2D) transition metal carbides and nitrides with chemical formula of $\mathbf{M_{n+1}X_nT_x}$, (where M= Sc, Ti, V, Cr, Zr, Nb, Mo, Hf, Ta; X is C and/or N, T is a surface termination unit such as hydroxyl (OH), oxygen(O) or fluorine(F) and n = 1, 2, or 3). MXenes have recently been synthesized through selective chemical etching of 3D MAX phases [5]. These two-dimensional structures are called as MXenes because they originate from the MAX phases by removing "A" elements and since they are structurally comparable to graphene [6,7]. Furthermore, the family of MXenes has been recently extended to ordered double transition metals carbides $M'_2M''C_2$ and $M'_2M''_2C_3$ [8].

MXenes are researched both theoretically and experimentally based on their unique attributes. Different experimental applications of MXene include transparent conductors [9], field effect transistors [10], supercapacitors [11], Li ion batteries [12], electromagnetic interface shields [13], fillers in polymeric composites [14], hybrid nanocomposites [15], purifiers [16] and many others. Theoretically, MXenes have been anticipated for many practical applications in electronic [17,18], magnetic [19,20], optical [21], thermoelectric [22], and sensing devices [23] along with new probable materials for catalytic and photocatalytic reactions [24] and nanoscale superconductivity [25]. 2D MXene, depending on the transition metal, bad gap, thermal stability, and surface functionality exhibits quite different electronic, magnetic, optical, and electrochemical properties that are rarely seen in their host compound i.e., MAX phase. Although a significant amount of research is based on these properties, but its magnetic properties are still unexplored.

Magnetism in MXene is an appealing topic for research and known for its application in spintronics. Two-dimensional layers of the MXene family exhibit a rich variety of magnetic properties and some of them are found to be ferromagnetic and anti-ferromagnetic in behaviour. The magnetism in MXenes can result from intrinsic properties of the constitutive transition metal, defects in monolayers, surface functionalities and through the synthetic procedures. It has been discovered and studied that proper surface functionalization turns MXenes to ferromagnets. Though some bare MXenes are predicted to be ferromagnets but experimentally most of them are usually terminated with F, O, OH or other atoms [26] since it is difficult to synthesize pristine MXenes.

Recently the swift technology development and computational methods have offered robust tools for understanding structures and properties of materials at the atomic scale, whereas first-principles simulations based on density functional theory (DFT) have shown us the powerful ability to predict and understand the mechanisms of phase stability, layered structures, and unusual properties of the MAX phases i.e., the host material. This has also proved to be true for MXenes: some significant issues have been simplified using DFT simulations one of them is that DFT simulations had played a vital role in the structures, properties, and potential applications of MXenes. In fact, the first recognized multilayer MXene structure, terminated $Ti_3C_2T_2$ layers were proposed with the support of a DFT modelling which helped researchers to get some more understandings into MXenes as metal ion batteries and other applications.

Khazaei et al has described MXenes electrical properties and applications briefly keeping in consideration the density functional theory (DFT) calculations [27]. Fatheema et al. has reported experimental and computational magnetic phase calculations that indicates the switching from superconductive-diamagnetic behaviour to ferromagnetism in La-doped $Nb_2C$-O-F MXene [28]. Babar et al. has also stated the diamagnetic behaviour of successfully synthesized 2D-$Nb_2C$ MXene and observance of superconductivity in the material [29]. Furthermore, the theoretical simulations based on DFT calculations confirmed that adsorption of and external atoms can generate localized moments in various representative two-dimensional (2D) crystals such as graphene, InSe monolayers and $MoS_2$, or even the long-range magnetic orders [30]. Gao et. al has reported about the functionalization $Ti_3C_2$ MXene by the adsorption or substitution of single metal atom using the density functional calculations [31]. Therefore, through density functional theory (DFT) structural, electronic, and magnetic properties of MXenes had been studied by taking different approximation methods. At present time, it is considered that the intrinsic magnetism of MXenes is a vital step in advancing their application. This work focuses on the magnetism of two compounds pristine $Nb_2C$-OF MXene and Ni-ad-$Nb_2C$-OF MXene. Theoretical exploration via first-principles calculations was conducted to study the electronic structure and magnetism of and Ni-ad-$Nb_2C$-OF MXene focusing on the effects of different functional groups on the electronic and magnetic properties of this material.

Furthermore, the effects of adsorption with transition metal atom in pristine $Nb_2C$-OF MXene on the properties were also studied. The close matching of ionic radii and oxidation states between Nb and Ni results stable configuration without significant lattice distortion highlight the importance of the present work. The ferromagnetic nature of Ni increases the electrical properties and magnetic of the $Nb_2C$-OF MXene, making it a suitable candidate for electronic and spintronic applications. Additionally, tuning the electronic structure can create active surface sites, thereby improving the catalytic activity of $Nb_2C$-O-F MXene for novel applications in energy storage, sensing, and catalysis.

## Sample Synthesis

Two compounds were prepared experimentally. Two dimensional (2D) $Nb_2C$ MXene sheets were prepared from 3D MAX Phase precursor (i.e., $Nb_2AlC$). After successful selective etching, Ni was deposited at $Nb_2C$-O-F surface.

**$Nb_2C$-OF Preparation.** Two dimensional MXene was prepared from $Nb_2AlC$ MAX phase precursor through selective wet chemical etching method. For successful etching results, 1 g of $Nb_2AlC$ MAX powder was immersed in 100 ml of HF (50% wt.) at ratio of 1:10 in Teflon beaker. The solution was constantly stirred for 40 hrs by Teflon-coated magnetic stirrer [32]. The temperature was optimized to be at 55 °C. The resulting deposits were washed many times with deionized water through centrifugation process for 5 minutes at 3500 rpm until the pH of supernatants was achieved to be ~6. The HF eliminates the Al layer, which, in sequence is replaced by F, O and/or OH. Every time, the supernatant was separated, and the powder settled at the bottom was removed from centrifuge bottles using ethanol. The resulting MXene powder was dried in vacuum oven at 40°C for 24 hrs. The resulting MXene was achieved in full amount without major losses.

**Ni-$Nb_2C$-OF Preparation.**

Ni doped $Nb_2C$ MXene was prepared by a straightforward hydrothermal method [33]. The $Nb_2CT_x$ powder was mixed in 30 ml deionized water by magnetic stirring for half-hour. The precursor solution was made by dissolving the crystals of nickel nitrate hexa-hydrate Ni (NO3)2.6H2O in 20 ml deionized water. The two prepared solutions were mixed, then magnetically stirred for 20 minutes. During the magnetic stirring, 50 % ammonia was added until pH was achieved to be 9. Hence the final solution was then shifted to a Teflon lined autoclave made of stainless steel (70ml) at 90°C for 16 hrs. Later, the reaction mixture was cooled at room temperature and the resultant precipitates formed were washed by deionized water and dried in convection oven at 60°C for 24 hrs. At the end, the dried product was stored in a glass vial and collected for further characterization.

For structural analysis of both the compounds and to observe the presence of material deposited on $Nb_2C$-O-F, X-Ray diffraction (XRD) was carried out by using Bruker D8 Advance system.

## Computational Framework

Before analysing the results of synthesized MXenes, it is worth reviewing computational methods used in the analysis. Amongst different computational methods, density functional theory (DFT) has proved to be a reliable method to predict various properties of synthesized material at an atomic level.

Hence all the first-principles calculations were performed within the framework of the density functional theory (DFT) based on the full potential linear augmented plane wave (FP-LAPW) potential. method implemented in Wien2K code [34]. The Perdew–Burke–Emzerhof (PBE) exchange–correlation function was used [35]. This is the most often used exchange-correlation functional for accurate and efficient calculations and is widely used means to study details of structures. The PBE-GGA approximation method at some point fails to calculate the correct results. Typically, in GGA the exchange-correlation energy does not capture the correlations effectively, therefore the so-called GGA+U method [36,37] (where U is Hubbard potential) is applied to correctly obtain magnetic order when the electron correlation in transition metals is important. The DFT analysis of the structural, ~~electrical~~ electronic, and magnetic properties for pristine $Nb_2C$-OF and Ni-ad-$Nb_2C$-OF was carried using Wein2k ~~software~~ code. These optimized compound structures were analysed for the different properties.

The internal geometry of pristine $Nb_2C$-OF and Ni-ad-$Nb_2C$-OF supercell 4×4×1 was optimized using 54 k-points in irreducible Brillouin zone (IBZ) distributed as 6×6×3 Monk horst–Pack grid. The wave functions in the interstitial regions are expanded in plane waves, with the plane wave cut-off chosen as $R_{MT} K_{max}$ 7.0 (here $R_{MT}$ represents the smallest atomic sphere radius and $K_{max}$ is the magnitude of the largest wave vector). The $R_{MT}$ radii are 1.92 a.u. for Nb atoms, 1.78 a.u. for O atoms, 1.82 a.u. for F atoms, 1.73 a.u. for C and 2.11 a. u for Ni atom. The spherical wave functions were expanded in spherical harmonics up to the maximum angular momentum $l_{max}$ = 10.

The pristine and Ni-ad-$Nb_2C$-OF MXene were optimized such that the forces on each atom were smaller than 1 mRy/a.u and the convergence criteria for the entire self-consistent calculations (scf) is such that the total energies do not exceed $10^{-5}$ Ry for successive steps. Moreover, the Hubbard “U” correction was also employed within the rotationally invariant ~~DFT~~ GGA + U approach for calculating its correct magnetic moment. After scf convergence the electronic band structure, density ~~od~~ of states and magnetic moment were calculated.

## Results and Discussion

### Characterization

The XRD patterns of powdered samples for pristine $Nb_2C$-OF MXene and 10% Ni doped $Nb_2C$-OF is shown in fig.1. By comparison of both compounds, the diffraction patterns clearly indicates that the spectrum of Ni-adsorbed $Nb_2C$-OF MXene is slightly different as compared to pristine $Nb_2C$-OF MXene spectrum. Within the range of 30° to 60°, the peaks for Ni adsorbed MXene are observed at 33.4° and 59.5° which agrees with the diffraction pattern of pristine $Nb_2C$-OF MXene. Here the decrease of peaks however indicates decrease in crystallinity of the structure with increase of Nickel concentration

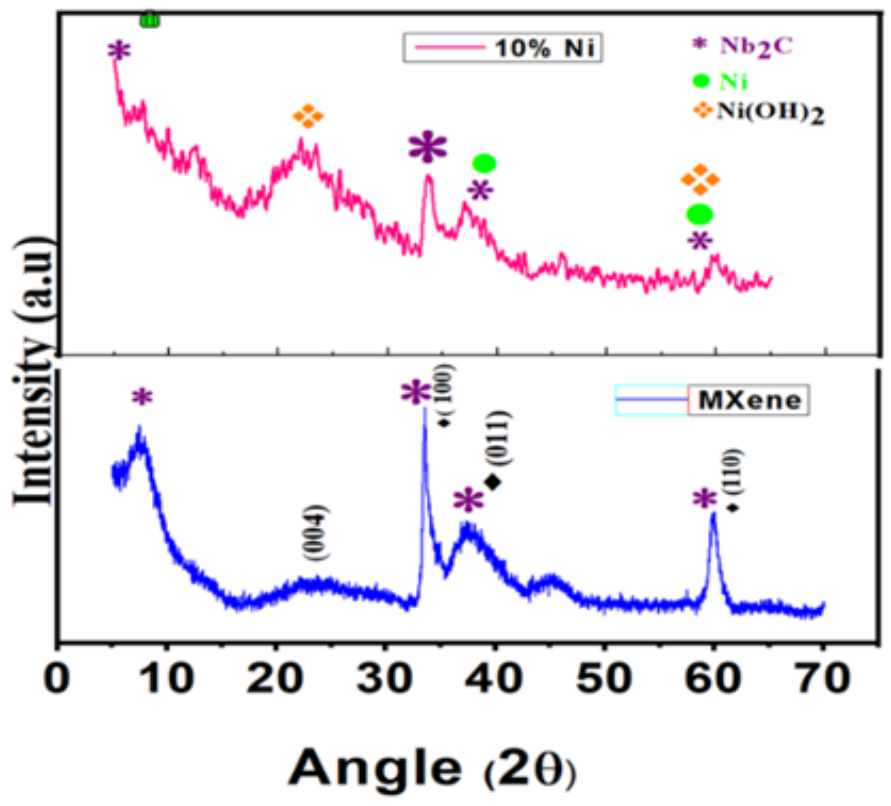

**Fig.1** XRD pattern for pristine $Nb_2C$-OF and Ni-ad-$Nb_2C$-OF

and results with disappearing of diffraction peaks shown for parent compound i.e., pristine $Nb_2C$-OF. The changed pattern is observed at 23° at plane (004) which indicates the presence of Ni-$(OH)_2$ oxides. For $Nb^{2+}$, the ionic radius is 0.07nm whereas for the dopant $Ni^{2+}$ the ionic radius is 0.069 nm. The ionic radius of $Ni^{2+}$ is in close approximation to that of $Nb^{2+}$, which indicates that the variations in diffraction pattern agree with the size of crystallite. The decline in the size of crystallite is due to the disturbance in the host $Nb_2C$-OF MXene lattice by the presence of $Ni^{2+}$ ions. The replacement of $Ni^{2+}$ ions in the interstitial position of host MXene lattice sites would disturb the intensity of the interstitial sites of pristine $Nb_2C$-OF which causes the peaks broadening [38]. The excuse of adsorption of $Ni^{2+}$ ions on the surface of pristine $Nb_2C$-OF is because of MXene negative potential surface caused by the presence of negative surface terminations (–O, –OH, –F). Therefore, pristine $Nb_2C$-OF MXene has a great capability to adsorb positively charged $Ni^{2+}$ ions. Ion exchange phenomenon may also take place at the surface between the $Ni^{2+}$ ions and negative surface terminations which results in formation of Ni-$(OH)_2$ oxides as shown in the XRD pattern [39,40,41].

## Computational Analysis

### Electronic Structure

To determine the stable structural ground state properties of pristine and Ni-ad-$Nb_2C$-OF MXene compound, the structural optimization was performed by minimizing the total energy with respect to the volume of the unit cell.

Originally, the experimental lattice parameters of used for the structural optimization process. Hence using the lattice parameters from the experiment, the pristine $Nb_2C$-OF structure is simulated by a supercell of slabs. Pristine $Nb_2C$-OF structure is designed consisting of 4 × 4 × 1 supercell. Each slab in supercell comprises of five layers. The C layer is introduced between Nb layers whereas Oxygen and Fluorine atoms are inserted to the system as functional group to terminate the surface (Fig.2). It's a layered hexagonal structure with the space group $P6_3/mmc$. The unit cell of $Nb_2C$ structure is extended into 4×4×1 supercell in ordered get the required concentration of Ni along with O/F functional group. The lattice parameters of 4×4×1 supercell are a=b = 12.47363 Å and c =22.59998 Å and α = β = 90° and γ = 120° which consists of 32 Nb atoms, 16 atoms of Carbon, 23 Oxygen atoms and 9 Fluorine atoms. Within the supercell, these slabs were separated by vacuum of about 18Å.

To understand the structure stability of Ni-$Nb_2C$-OF MXene, two structures of 2×2×1 supercell for Ni-Nb2C-OF MXene were optimized (see Fig. 3) in such a way in first case Nb atom was replaced with Ni atom at suitable position while in second case Ni atom was allowed to adsorb on the surface of structure forming bonds with the surface functionalization. The calculated ground state energy for adsorbed and doped structure calculated was -65777.5818 Ry and -58136.5490 Ry respectively.

Since more negative the minimization energy more stable is the structure. Hence the adsorbed structure was found more stable than the doped one. For further electronic structure calculation, the 4 × 4 × 1 supercell were used for pristine and Ni adsorbed Nb2C-OF MXene as shown in shown in Fig. 4.

### Electronic Properties

Figure 5 illustrates the electronic band structure for Ni-ad-$Nb_2C$-OF MXene. The band structure of Ni-ad-$Nb_2C$-OF is calculated along high-symmetry directions of the irreducible Brillouin zones (IBZ) (Y-Γ-X -Y) respectively while the fermi level was set at 0 eV. Most of the functionalized MXenes are found metallic in nature. As clearly seen from fig 5, the electronic band structure calculations reveal that Ni-ad-$Nb_2C$-OF MXene is metallic in nature as there exists no band gap [42]. The fermi energy is all crowded as different energy levels are seen overlapping whereas the conduction band is much occupied and crowded with energy states as compared to the valence band. Many of MXenes contain heavy 4d and 5d transition metals which significantly affects the electronic structures. Fig.6 shows the total density of states (TDOS) versus energy (eV) plot for pristine $Nb_2C$–OF and Ni ad-$Nb_2C$–OF. The non-overlapping states above the fermi level as seen in band structure are visible as clear separate peaks in TDOS plot for Ni-ad-$Nb_2C$-OF MXene. While observing Nickel in TDOS plot, we can see its contribution throughout the valence and conduction band with more involvement near fermi level. Since there is no band gap in the plot as the states are spreading in narrow and wider distribution and overlapping near fermi level. Hence metallic behaviour of Ni-ad $Nb_2C$-OF MXene can be clearly proved from this TDOS plot. Since Ni and Nb are both transition metals with partially filled d orbitals, therefore the projected band structure calculation reveals that the bands near the Femi energy originate from d orbitals of transition metals. Their magnetic nature is predicted to be based upon the magnetic transition metal elements (e.g., Cr, Mn, V, Fe, and Ni), doping or adsorption of magnetic atom or near neighbour effect of interface [43]. In conclusion significant amount of energy states observed in the vicinity of fermi level which confirms the metallic nature and therefore, the fermi energy is located at the d-bands of transition metal Niobium and Nickel.

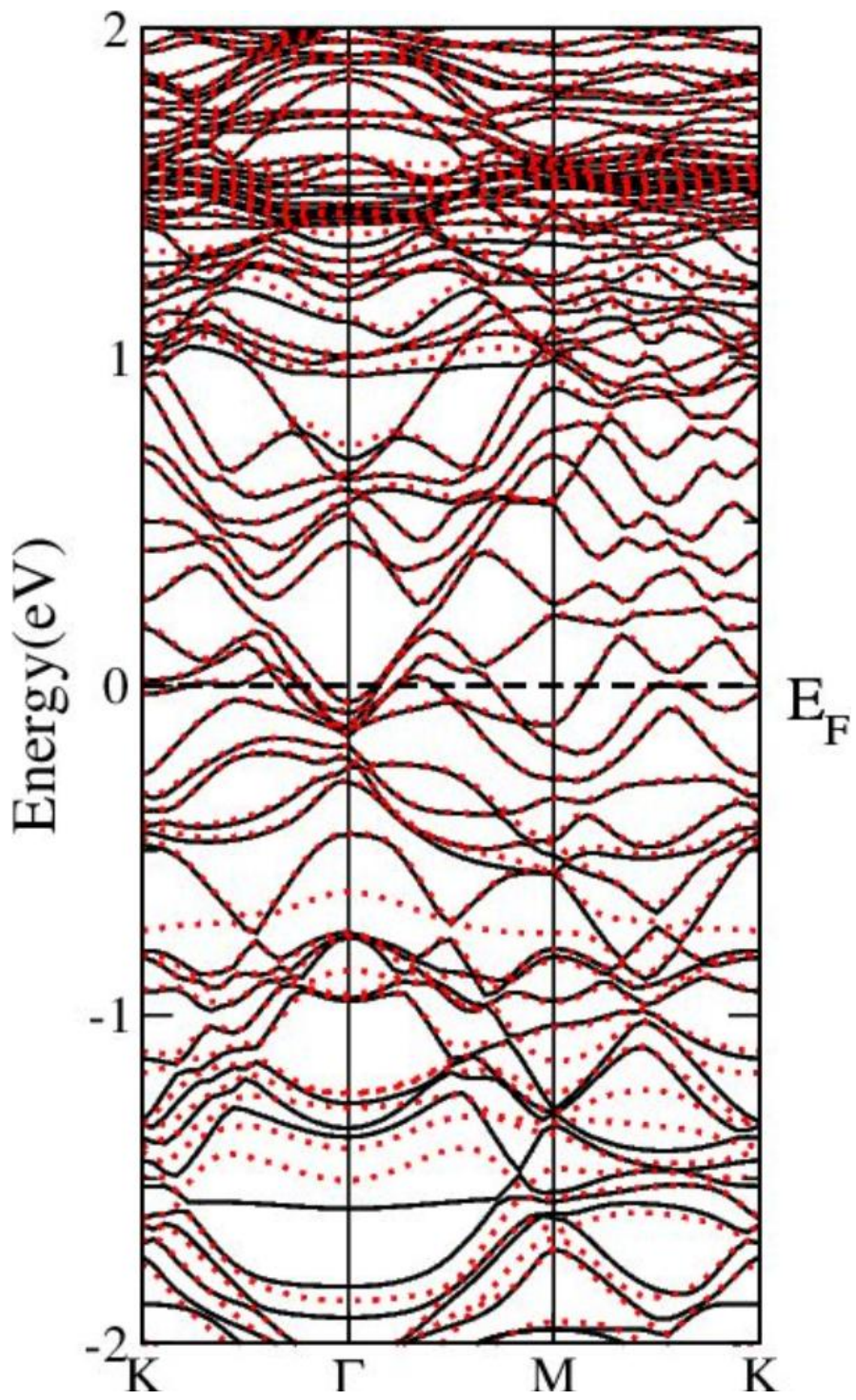

**Fig.5** Band Structure displaying metallic behaviour with no band gap.

## Magnetism

Magnetism in Ni-ad-$Nb_2C$-OF MXene has not been studied and researched till now where as pristine $Nb_2C$-OF MXene has been predicted to be diamagnetic in recent publication [44]. The total

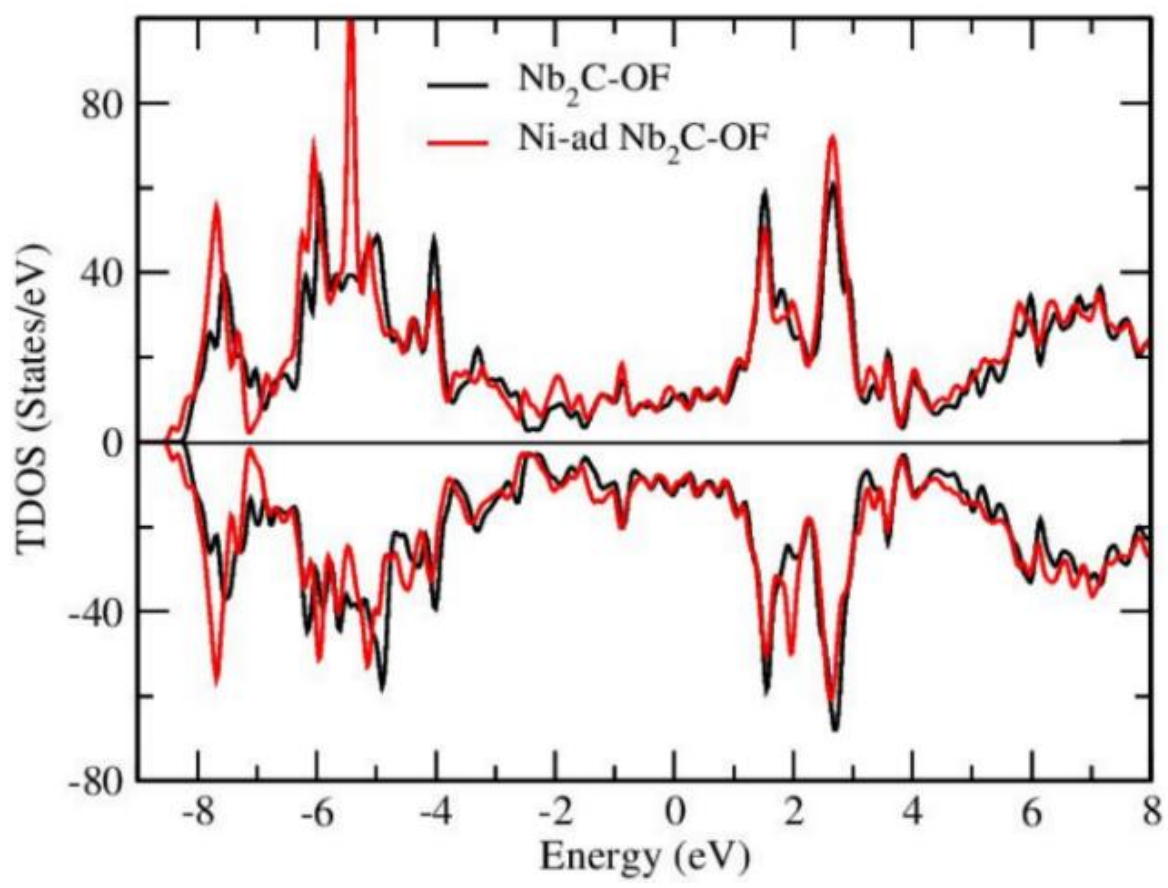

**Fig.6** TDOS/eV plot for $Nb_2C$-OF and Ni-ad-$Nb_2C$-OF

magnetic moment of pristine $Nb_2C$-OF MXene calculated in our case is -0.02734 $\mu_\beta$. The value is small yet is important as it confirms the diamagnetic behaviour according to the literature. The results confirmed that negative magnetic moment is mainly contributed entirely by the Nb-atoms. (As seen from plot of magnetic moment mentioned in Supplementary info S1).

Nickel is one of the elements that is ferromagnet at room temperature. When Ni is adsorbed on the pristine $Nb_2C$-OF MXene, it can form bonds with functional groups $F^-$ and $O^-$ atoms on the surface, but it was observed for stable structure, that only oxygen and niobium atoms form bond with nickel atom. (Fig 4)

**Density of states (DOS) and Partial density of states (PDOS)**

From Fig.6, its visible that TDOS/eV for $Nb_2C$–OF has a continuous spread all over the region from valence band to conduction band with enough peak height. With the adsorption of certain foreign material in pristine $Nb_2C$–OF, the system gives same continuous spread all over the region with the highest peak of 90 states/eV. To know more about the metallic behaviour and the adsorption, Ni-adsorbed $Nb_2C$-OF total density of states (TDOS) and projected density of states (PDOS) calculated for detail analysis using PBE-GGA+U approximation are presented in fig.7 and fig.8.

The total density of states (TDOS) versus eV plot for the Ni-ad-$Nb_2C$-OF MXene is shown by fig.7. The plot was computed by using the spin-GGA+U where U= 6eV using 150 k-points with the k-mesh 6 × 3 × 3. The TDOS/eV plot for spin up and spin down DOS seems not to be mirror of each other. There is a reasonable energy difference between spin up and spin down states of adsorbed compound which indicates its ferromagnetic nature. As seen from the TDOS plot, the peaks for all elements Ni, Nb, C, O, F are clearly visible indicating the contribution of each element in this compound. The TDOS peak demonstrates that all atoms in super cell are strongly bonded and gives the highest peak within the range of -7eV to -6 eV While observing Nickel in TDOS plot, we can see its contribution throughout the valence and conduction band with more involvement near fermi level. Since there is no band gap in the plot as the states are spreading in narrow and wider distribution and overlapping near fermi level. Hence metallic behaviour of Ni-ad $Nb_2C$-OF MXene can be clearly proved from this TDOS plot.

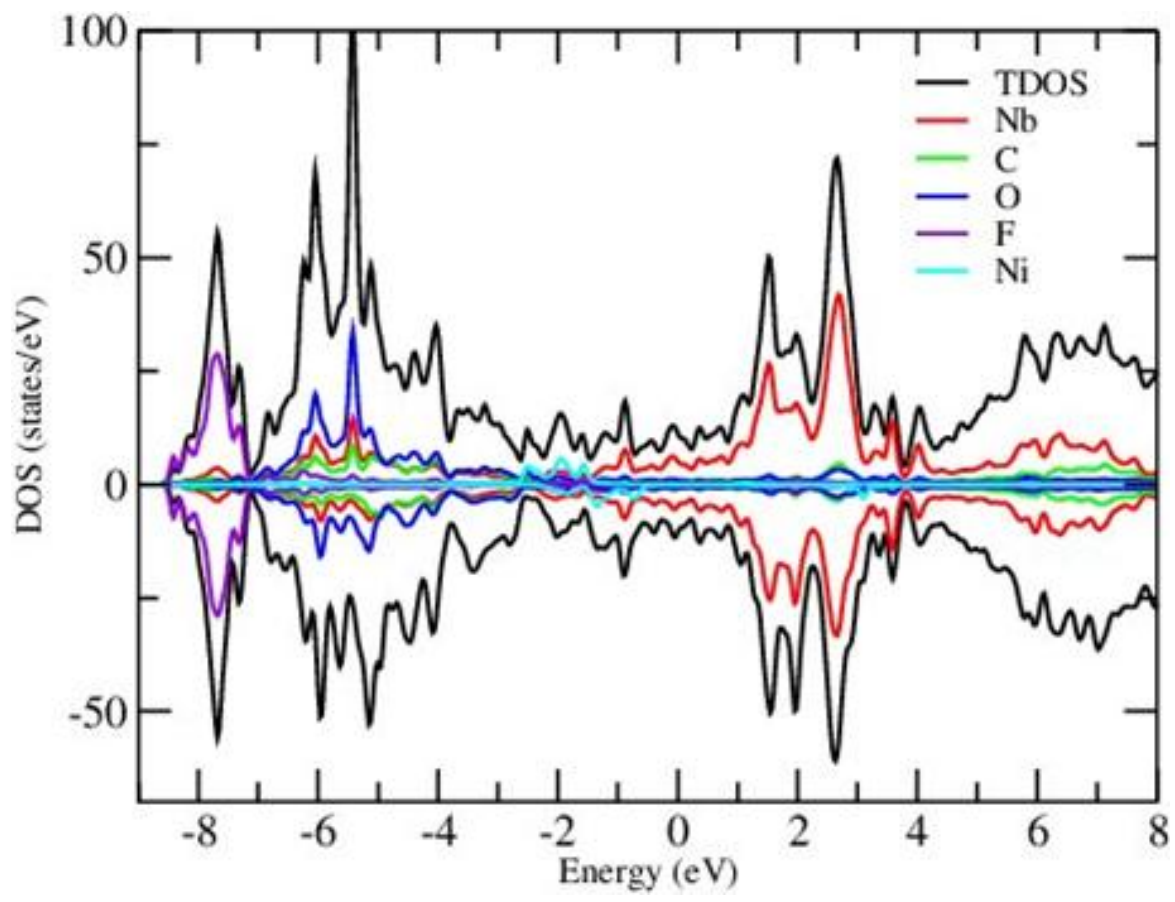

**Fig.7** Detailed TDOS plot for Ni-Adsorbed $Nb_2C$-OF MXene.

**Partial density of states (PDOS)**

As a notable case to investigate the involvement of each atomic orbital in TDOS, the projected density of states (PDOS) for of Ni-ad-$Nb_2C$-OF MXene are decomposed as angular momentum channels are displayed as PDOS versus energy plots (Fig. 8a-e). From the fig. 8(a) Nb d-orbital has the highest spin up and spin down peaks as compared to Nb-s and Nb-p orbital. The Nb-d orbital has a maximum peak of 1.2 eV within the conduction band and small contribution all around the fermi level. Whereas while analysing the Nickel PDOS/eV plot, Ni d-orbital (figure 8b) has the highest spin up and spin down peaks within the range of -2 eV to 2 eV i.e., around the fermi level with a small contribution of s-orbital. The Nb-d orbital contribution is suppressed by the high peaks for Ni atom. Hence the partially filled atomic orbitals of Ni (d-orbital) and Nb(d-orbital) are overlapping and resulting in hybridization [45]. Therefore, the states all around the fermi level result mainly from the hybridization between Nb and Ni d-orbitals crowding the fermi level. By comparison to the PDOS of other elements of compound (i.e., carbon (C) and surface functionalization's $O^-$, $F^-$), small peaks can be seen around the fermi level for C and O with a very limited contribution from Fluorine (F). Upon $O^-$ and $F^-$ functionalization, only C-p orbital of the attached chemical groups are also hybridized with Nb and Ni d-orbitals. While comparing the PDOS's (fig.8(b-e)) it is seen that the hybridization between Ni-d, C-p and O-p orbitals is noticeable with a very less contribution from F-p orbital. Fig. 9(i-iii) PDOS/eV plot for atoms in Ni-ad-$Nb_2C$-OF having significant contribution in hybridization. The figure clearly indicates a major contribution from Nb atoms (i.e., Nb1 and Nb9 along with Nb4 and Nb8) with a negligible contribution around the fermi level from F atoms. This means that the metallic nature resulting from hybridization is contributed by specific Nb atoms. More precisely, it can be concluded that Nb-d orbital contribution is suppressed by the high peaks for Ni atom and the three orbitals (i.e., Ni-d, Nb-d, and C-p along with O-p contribute significantly to hybridization. Consequently, several new states are generated above and below the fermi level crowding it completely resulting in zero band gap.

In this work, Ni was adsorbed on pristine $Nb_2C$-OF MXene which resulted in crowding the fermi level and increasing the number of DOS crowding the fermi level. Moreover, Ni adsorption increased TDOS, specifically around the fermi level appearing within the range of -2 eV to 0 eV with maximum peak of 5 DOS/eV. Moreover, the spin up and down peaks are totally different showing its magnetic behaviour. In conclusion, this hybridization results in zero electronic band gap and metallic nature of the compound. In fig.8, the region from -4eV to 4 eV demonstrates hybridization among all atomic orbitals with the most dominated hybridization near fermi level. This dominated hybridization results in the magnetic and semi-metallic nature of Ni-ad-$Nb_2C$-OF MXene. Hence, zero band gap and metallic property of this compound is proved from this PDOS versus eV plot and is also supported by our electronic band structure. Hence a small contribution of Ni atoms has dominated the major dominated contribution of Nb atoms in Ni-$Nb_2C$-OF MXene. Sharma Y. and Srivastava P [46] have also discussed about the reduction of band gap in nickel-doped arsenic triselenide ($As_2Se_3$) caused by contributions of Ni-d orbitals in the conduction band minimum near fermi level. Similarly, high peaks of Ni-d orbitals in both valence and conduction band with minimum Ni concentration in Ni-doped cadmium sulphide (CdS) has also been reported [47].

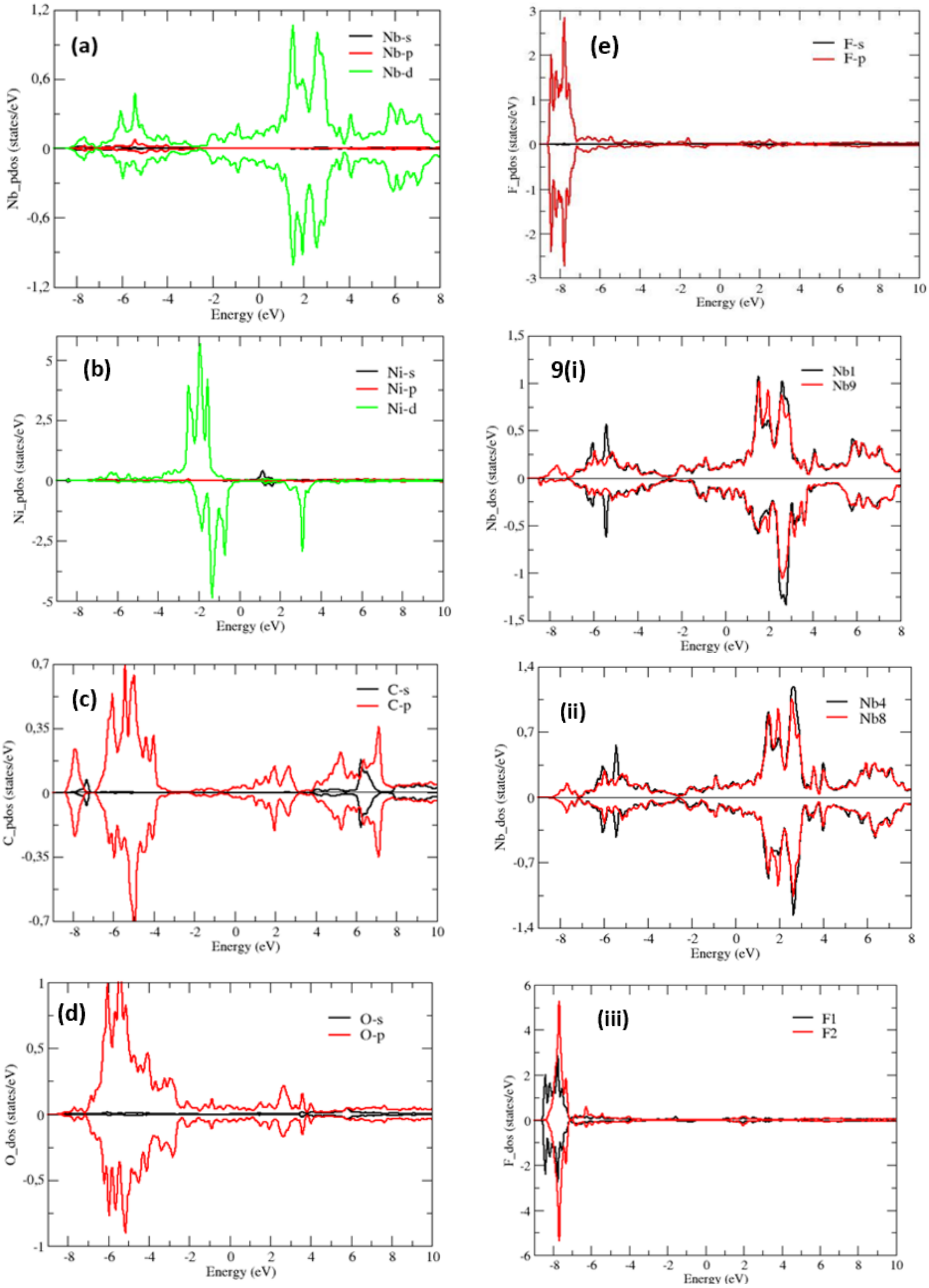

**Fig. 8(a-e)** PDOS/eV plot for Ni-ad-$Nb_2C$-OF atoms revealing contribution of s, p, d orbitals. **Fig. 9(i-ii)** PDOS/eV for specific Nb atoms contributing significantly to magnetic analysis in Ni-ad-$Nb_2C$-OF. **Fig.9(iii)** PDOS/eV for specific F atoms demonstrating their contribution to magnetic analysis in Ni-ad-$Nb_2C$-OF.

As far as we know there is no detailed study on DOS of Ni-adsorbed $Nb_2C$-OF MXene, the present results suggest the possibility of reduction of electronic states near fermi level and giving no band gap. Hence, we expect future experiments to verify our results.

## Magnetic Moment

The magnetic moment for Ni-ad-$Nb_2C$—OF is found to be $+1.01516\mu_\beta$ a significant value showing ferromagnetic nature. The strong intensity for Ni-d orbitals (Fig.9b) in the vicinity of the fermi level is predicted to be the reason for magnetic moment with the ferromagnetic nature.

Fig.10 shows the plot of magnetic moment for all the atoms in 4×4×1 supercell for pristine and adsorbed structures. The pristine structure consists of 32Nb, 32 functional groups atoms i.e., $O^-$ and $F^-$, 16 C's whereas for adsorbed system an additional Ni atom is added to structure giving a total of 81 atoms. Considering Fig.4, optimized simulated structure of Ni-ad-$Nb_2C$-OF MXene, Ni atom forms a bond with two O and one Nb which highly influence its magnetic moment. The magnetic moment plot for adsorbed system is continuous spread with two main high peaks. The two main peaks are due to magnetism of oxygen bonded with Ni. For the analysis, the two oxygen atoms at position 36 and 42 shows high peak representing the large magnetic moment. The large magnetic moment of oxygen is induced from single Ni atom. The large magnetic moment contributed by Ni atom at position 81 shows clear high peak for positive magnetic moment. Hence Ni magnetic moment has induced the oxygen magnetic behaviour resulting in switching from diamagnetic behaviour ferromagnetic nature with larger magnetic moment for Ni-ad- $Nb_2C$—OF. While looking at the magnetic moment spectrum, along with two maximum high peaks three negative peaks are also under concern. The extreme negative peak in the range of 0 to 20 atoms shows negative magnetic moment for adsorbed compound. Within this range, Nb atoms are mostly contributing to the negative magnetic moment (as given in S1). From optimized Ni-ad $Nb_2C$-OF structure (fig.4), considering the side view, the atoms linked with adsorbed Ni atom are Nb1, Nb4, Nb8, and Nb9. As seen from the PDOS/eV (fig.9.i) for these atoms, Nb1 and Nb9 has a dominant contribution in DOS plot with some contribution around the fermi level. Likewise looking at Nb4 and Nb8 PDOS/eV (fig.9.ii) shows their contribution to magnetic moment. These Nb atoms contribute to the hybridization that affects the magnetic nature of Ni-ad-$Nb_2C$-OF. The second major negative peak between 50-60 atoms is due to the F1 and F2 atoms. PDOS/eV (fig.9iii) for F1 and F2 demonstrate that their minor contribution in conduction band with zero presence around the fermi level. The contribution from F1 and F2 can be seen limited to the valence band. While comparing the magnetic moment calculated for all individual atoms, the oxygen and niobium atoms forming bond with nickel were found to have more magnetic moment than other elements. (Table in S1).

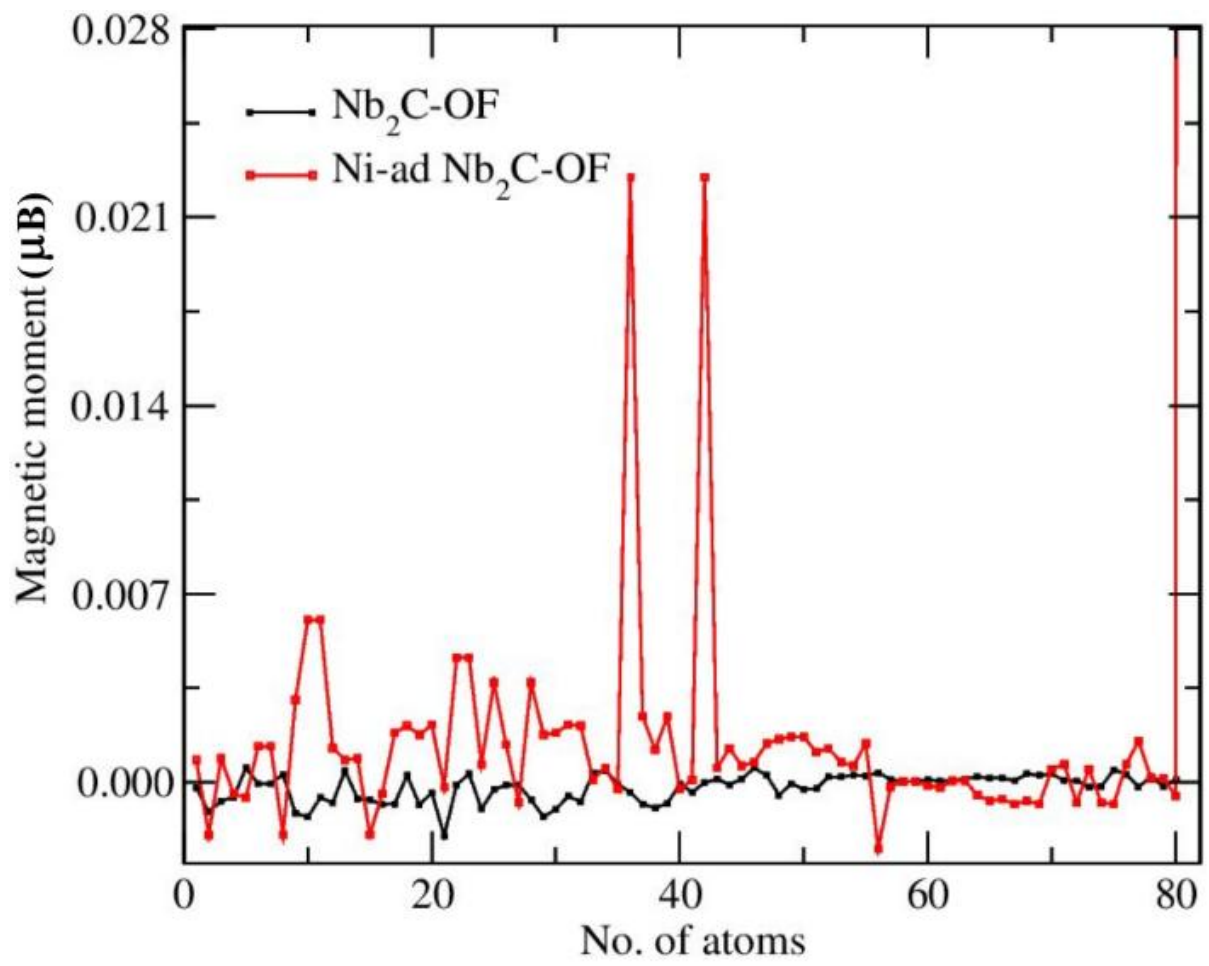

**Fig.9.** Magnetic moment for Pristine and Ni-ad-$Nb_2C$-O-F MXene atoms in 4×4×1 supercell where position 1-31 is Nb, 32-54 is O, 55-63 is F, 64-80 is C and last at 81 position is Ni.

The magnetic moment calculated for O atoms and Ni is positive with large value while the magnetic moment for Nb linked to these atoms is very less with negative value. The large positive magnetic moment dominates the negative magnetic moment of Nb. Hence the negative magnetic moment from the Nb atom is decreased and a positive magnetic moment is revealed showing the Ferromagnetic behaviour. In simple words, the switching of the diamagnetic behaviour ferromagnetic is reasoned due to the surface termination and the adsorption of Nickel atom.

## Conclusion

Structural, electronic, and magnetic properties of transition metal Ni-adsorbed $Nb_2C$-O-F MXene have been calculated for magnetic applications. The electronic structure analysis under density functional theory (DFT) calculations implemented in Wien2k confirms the metallic nature of Ni-$Nb_2C$-OF MXene. Moreover, theoretical simulations based on DFT calculations showed that absorption of foreign transition metal Nickel along with surface terminations changed the diamagnetic nature of pristine $Nb_2C$-OF too ferromagnetic nature. This theoretical study on Ni-adsorbed $Nb_2C$-O-F MXene compound has been conducted using PBE-spin-GGA and PBE spin-GGA+U approach which are employed within the framework of DFT using Wien2K code. The magnetic moment for Ni-ad- $Nb_2C$-O-F MXene is +1.01516 $\mu_\beta$. The magnetic result presented here is novel and first report on magnetic properties of Ni-ad- $Nb_2C$-O-F MXene. The present study and magnetic analysis open-up a tremendous possibility of their applications in spintronics as well super-spintronics. For future studies, performing the detailed further theoretical analysis into consideration might provide deeper insights about Ni-ad-$Nb_2C$-OF MXene which can include antiferromagnetic analysis and using other exchange correlational functionals as a tool for deeper DFT magnetic calculations.

## Acknowledgements

The authors are thankful to Higher Education Commission (HEC) of Pakistan for providing research funding under the Project No.: 20-14784/NRPU/R&D/HEC/2021. Saleem Ayaz Khan is grateful for support from QM4ST project financed by the Ministry of Education of the Czech Republic grant no. CZ.02.01.01/00/22_008/0004572, co-funded by the European Regional Development Fund.

## Authors Contribution

Zarah Khan has analysed experimental and computation data and has written the manuscript, Saleem Ayaz Khan has extensively helped in computational analysis and co-supervised the project, Ayesha Zaheer synthesized the sample and helped in a few characterizations, and Syed Rizwan conceived the research concept, wrote the manuscript, and supervised the complete project.

## Funding Sources

The authors thank the Higher Education Commission (HEC) of Pakistan for providing research funding under the Project No.: 20-14784/NRPU/R&D/HEC/2021 and QM4ST project financed by the Ministry of Education of the Czech Republic grant no. CZ.02.01.01/00/22_008/0004572, co-funded by the European Regional Development Fund.

## Conflict of Interest

There are no conflicts to declare.

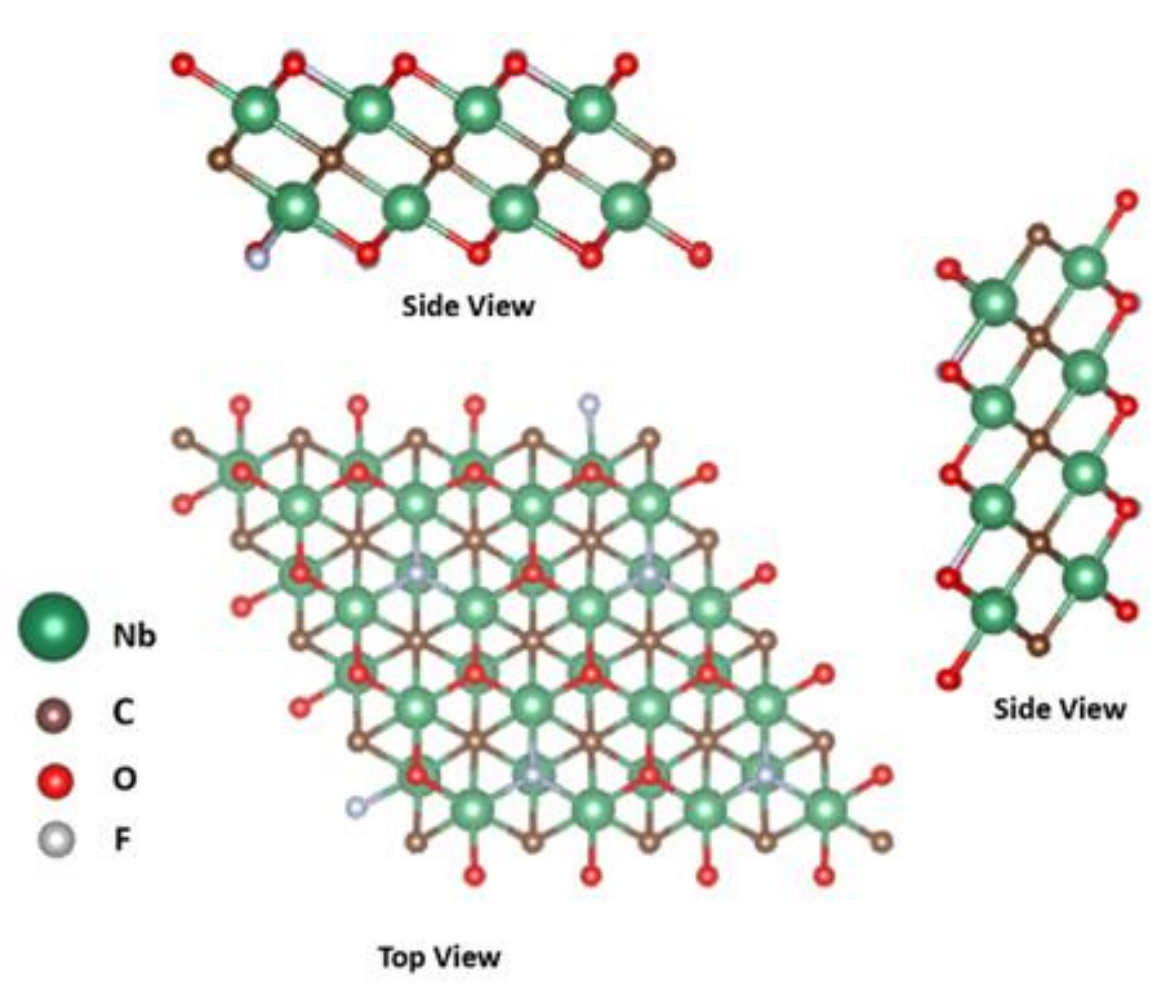

**Fig.2** $Nb_2C$-OF MXene structure terminated with $O^-$ and $F^-$ showing top and side views

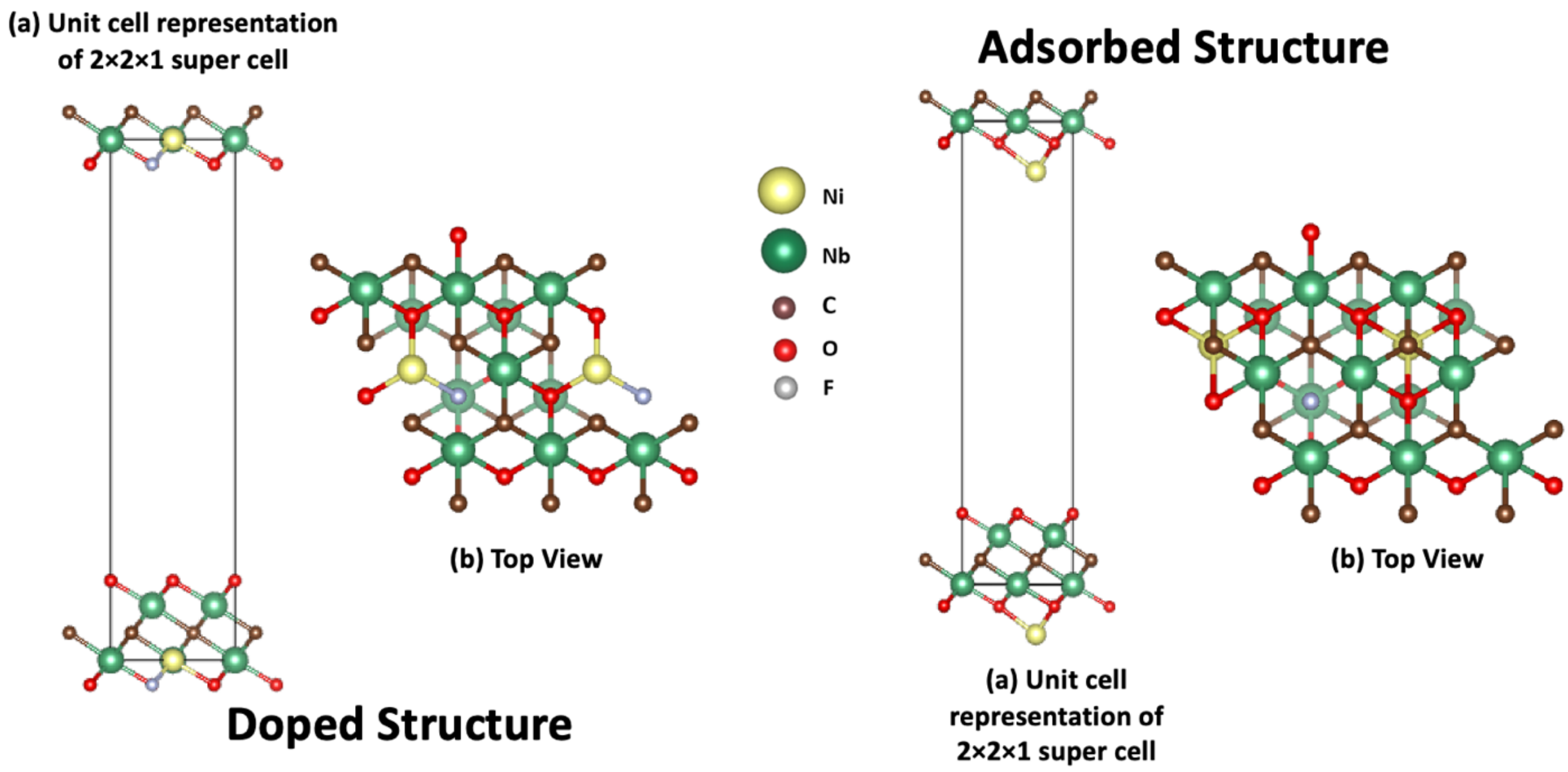

**Fig.3** 2×2×1 supercell of Doped and Adsorbed Ni-$Nb_2C$-OF MXene

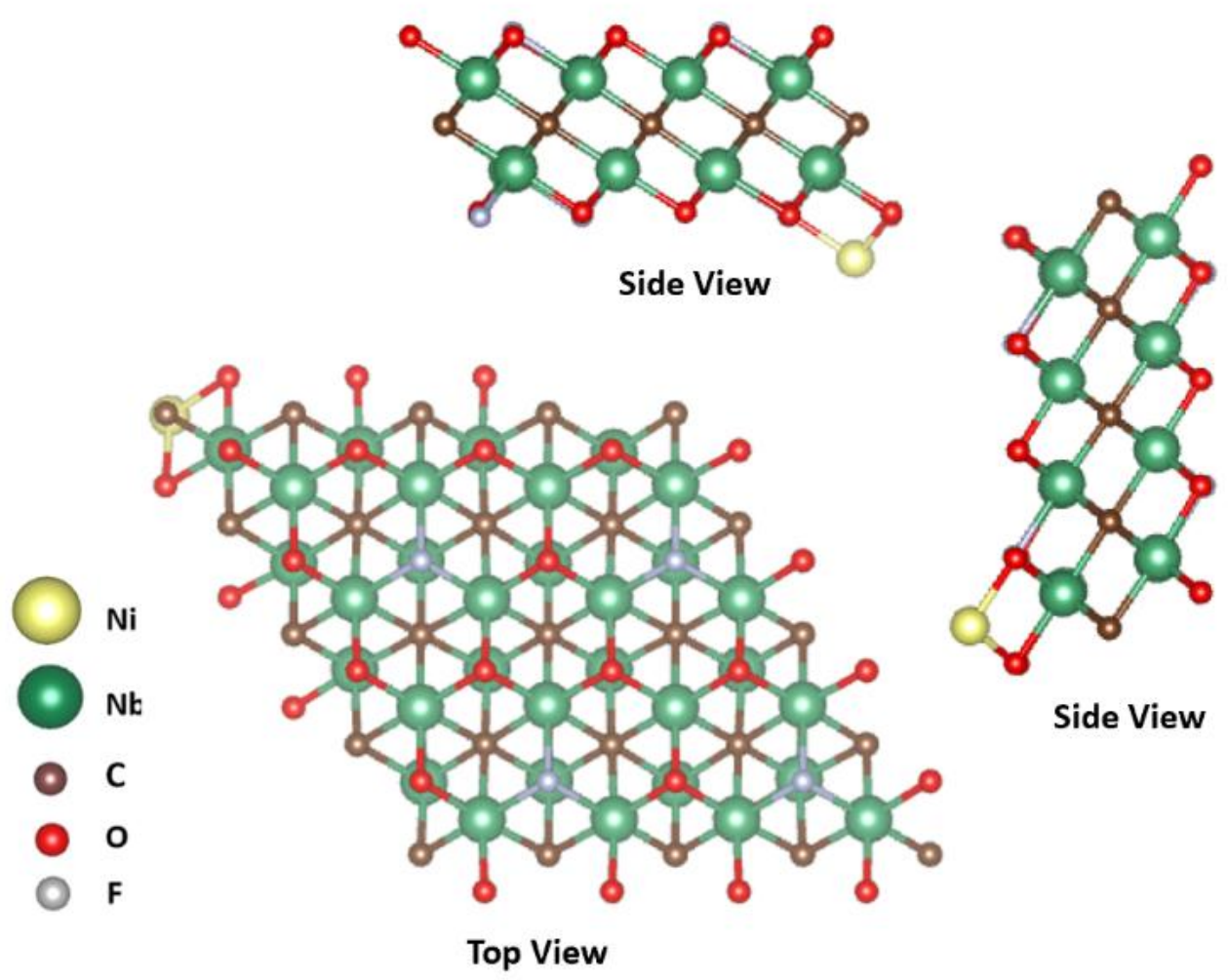

**Fig.4** Top and side views demonstrating 4 × 4 × 1 supercell of Ni-ad-$Nb_2C$-OF MXene structure.